\documentclass[12pt]{iopart}
\pdfoutput=1

\usepackage[T1]{fontenc}
\usepackage[utf8]{inputenc}

\usepackage{iopams}     
\usepackage{graphicx}
\usepackage{color}
\usepackage{cite}
\usepackage{soul}

\expandafter\let\csname equation*\endcsname\relax
\expandafter\let\csname endequation*\endcsname\relax
\usepackage{amsmath}

\usepackage[colorlinks=true]{hyperref} 

\begin{document}

\title[Diffusion with doubly stochastic resetting]{Diffusion with doubly stochastic resetting}

\author{Maxence Arutkin and Shlomi Reuveni}

\address{School of Chemistry, Center for the Physics \& Chemistry of Living Systems, Sackler Center for Computational Molecular \& Materials Science, Tel Aviv University, 6997801, Tel Aviv, Israel}
\ead{maxence@tauex.tau.ac.il,shlomire@tauex.tau.ac.il}

\begin{abstract}
Diffusion with stochastic resetting, instantaneous returns of a diffusing particle to a reference point, creates a stationary probability distribution. The paradigm is extended here to a doubly stochastic protocol in which the resetting rate itself fluctuates in time and relaxes on its own timescale. An exact steady-state solution reveals three spatial regimes: a fluctuation-dominated core near the origin, a power-law regime at intermediate distances, and a far-field exponential decay fixed by the rate-relaxation time. These results expose how the instantaneous rate, mean rate, and relaxation time come together to determine the non-equilibrium steady state.
\end{abstract}

\section{Introduction}
Stochastic resetting, where a system is randomly returned to a reference state, can create non‑equilibrium steady states (NESS) in dynamics that would otherwise be unbounded~\cite{evans2019resetting,gupta2021briefreview,talfriedman2020experimental}. A foundational result by Evans and Majumdar~\cite{evans2011diffusion} showed that diffusive motion with constant-rate resetting reaches a stationary probability distribution, offering a simple yet powerful mechanism to control systems far from equilibrium. A notable consequence is a dynamical phase transition in the relaxation toward stationarity~\cite{majumdar2015dynamical}. 

Constant‑rate resetting has proven useful in many contexts. Examples include optimal search processes~\cite{reuveni2016optimal,pal2020search}, enhanced chemical reaction rates~\cite{reuveni2014role,rotbart2015michaelis}, protein‑folding models~\cite{blumer2022stochastic}, and DNA‑targeting searches in confined biological environments~\cite{eliazar2007search,benichou2009search}. However, many real-world systems, such as enzyme kinetics, fluctuating search strategies, and noise-driven processes, rarely conform to constant-rate resetting and often exhibit fluctuations due to environmental or intrinsic factors. 

Recent studies have examined protocols with a time‑dependent resetting rate and uncovered rich behavior once the rate is no longer fixed. Notably, a power‑law distribution of reset intervals can erase the usual exponential tail and, for some exponents, remove the steady state entirely~\cite{nagar2016diffusion}. In addition, periodic reset protocols have been shown to surpass Poissonian resets in reducing search times under specific conditions~\cite{bhat2016stochastic,pal2017first}. 

Pal et al.~\cite{pal2016diffusion} investigated a resetting process where the resetting rate is a function of time elapsed since the last reset event. In their model, the rate is reinitialized, together with the particle, following every resetting event. This time‑dependent modulation produces a variety of long‑term behaviors that depend on the chosen rate function, and can be shown equivalent to a formulation where the times between resetting events are taken from an arbitrary distribution~\cite{pal2017first}.

Bodrova et al.~\cite{bodrova2019nonrenewal} investigated non-renewal resetting of scaled Brownian motion, in which each reset returns the particle to the origin while leaving the time-dependent diffusion coefficient $D(t)\propto t^{\alpha-1}$ unchanged. Their closed-form expressions for the MSD and the full propagator demonstrate that, for suitable $\alpha$ and waiting-time distributions, the MSD saturates even though the position distribution never reaches a NESS. A comparison with the renewal counterpart of the problem, in which both position and internal age are restarted at each reset~\cite{bodrova2019scaled}, underscores how retaining memory in stochastic resetting processes profoundly alters long-time relaxation and first-passage statistics.

\begin{figure}[t]
  \centering
  \includegraphics[width=\textwidth]{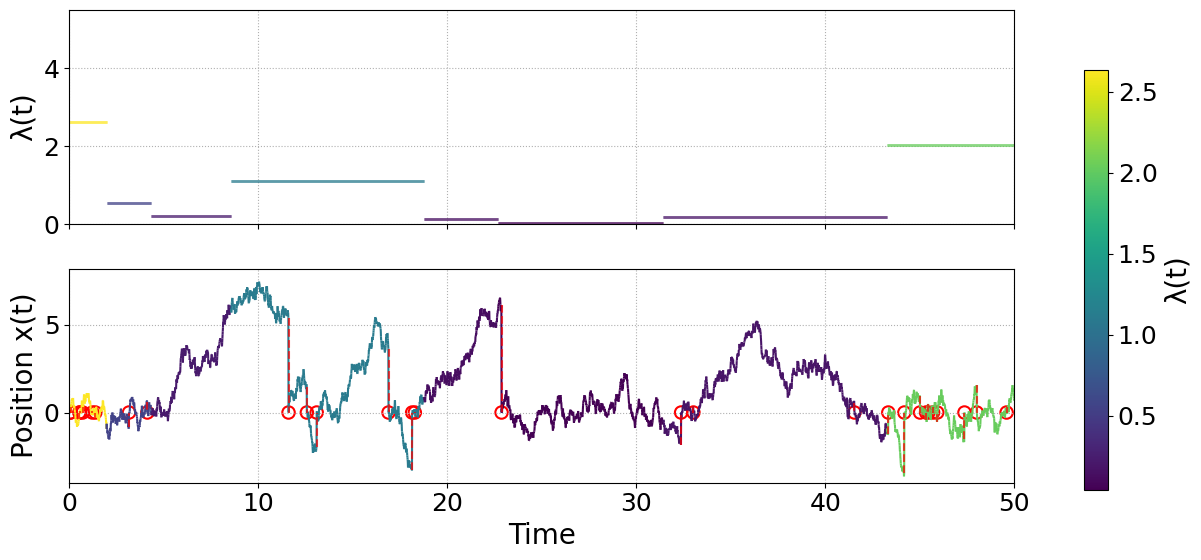}
  \caption{\textbf{Realization of diffusion with doubly stochastic resetting.} \emph{Top panel:} Time series of the resetting rate $\lambda_t$, held constant over independent and identically distributed successive intervals with exponentially distributed lengths $\tau_i\sim\mathrm{exp}(\tau_r^{-1})$ and independently redrawn from the distribution $f_\lambda(\lambda)$.  Each horizontal line represents one such interval; its color encodes the instantaneous value of $\lambda_t$, as indicated by the vertical color-bar on the right. \emph{Bottom panel:} A single diffusive trajectory $x_t$ subject to resetting at rate $\lambda_t$, colored by the same color as the instantaneous resetting rate.  Resetting events are highlighted by red dashed vertical lines, thereby emphasizing the impact of rate fluctuations on the timing of returns.}
  \label{fig:reshuffled_rate_trajectory}
\end{figure}

Here, we study a non-renewal resetting protocol in which the resetting rate evolves independently of the particle dynamics and restart events, remaining unchanged when particles return to the origin, allowing both its instantaneous value and long-time fluctuations to shape the steady state. As illustrated in Fig.~\ref{fig:reshuffled_rate_trajectory}, this doubly stochastic resetting induces alternating epochs of frequent returns and extended excursions, enabling the system to intermittently explore distant regions of space before being abruptly recalled to the origin. 

We investigate a doubly stochastic resetting diffusion model by deriving the position probability density function (PDF) and analyzing the resulting NESS. The protocol therefore belongs to the class of non‑renewal resetting because the internal variable that triggers resets retains its memory and continues to evolve between events. Our approach encompasses both time-dependent and stochastic resetting rates, allowing us to identify distinct regimes within the NESS based on the interplay between diffusive dynamics, stochastic resetting, and the temporal behavior of the resetting rate itself. 

We find that the stochastic resetting rate organizes the steady state into a super‑statistical core and a self‑averaged tail, controlled by the rate relaxation time $\tau_r$. At short distances, the system exhibits super-statistical behavior, where fluctuations in the resetting rate dominate. For long distances, the fluctuations in the resetting rate average out, and the PDF converges to a form governed by an effective mean rate. However, depending on the rate relaxation time $\tau_r$, this effective resetting rate may differ significantly from the typical resetting time $\bar\lambda^{-1}$. When  $\tau_r$  is small, a rapid relaxation of the rate results in a self-averaging effect, and the NESS approaches that of a constant resetting rate $\bar\lambda^{-1}$. Conversely, when  $\tau_r$ is large, the effective resetting rate is considerably lower than the mean rate due to a survival bias toward lower resetting rates for particles which traveled over long distances. This effect introduces a significant deviation in the tail, highlighting the impact of the resetting rate's temporal variability on the NESS structure. By quantifying these impacts, we provide a comprehensive view of the NESS under variable, history-dependent resetting protocols, demonstrating how the relaxation dynamics of the resetting rate shape the system’s steady-state properties in both local and asymptotic regimes.

\section{Diffusion equation with a fluctuating resetting rate}
We investigate the propagator $ P_r(x,t) $ for a diffusion process subject to a time-dependent and stochastic resetting rate $ \lambda_t $ to the origin. The diffusion with resetting is governed by the modified diffusion equation:
\begin{equation}
\frac{\partial P_r(x, t)}{\partial t} = D\, \frac{\partial^2 P_r(x, t)}{\partial x^2} - \lambda_t\, P_r(x, t) + \lambda_t\, \delta(x),
\end{equation}
where $ D $ is the diffusion coefficient, and $ \delta(x) $ represents the instantaneous resetting to the origin at position $ x = 0 $. In this setup, the resetting rate $ \lambda_t $ varies over time in a stochastic manner, adding complexity to the diffusion dynamics. Our aim is to derive an expression for $ P_r(x, t) $ that fully accounts for both the time dependence and stochastic nature of $ \lambda_t $.

We begin by applying the Fourier transform of the spatial coordinate. The resulting equation in Fourier space reads:
\begin{equation}
\frac{d \hat P_r(k,t)}{d t} = -Dk^2\, \hat P_r(k,t) - \lambda_t\, \hat P_r(k,t) + \lambda_t.
\end{equation}
We can rewrite the equation as:
\begin{equation}
\frac{d\hat P_r(k,t)}{dt} + [Dk^2 + \lambda_t]\, \hat P_r(k,t) = \lambda_t.
\label{Eq:ODEFourier}
\end{equation}
This linear ordinary differential equation governs the time evolution of the propagator in Fourier space, accounting for both diffusion and the effect of stochastic resetting. Assuming the particle initially starts at the origin, $x_0 = 0 $, the initial condition simplifies to $\hat P_r(k, 0) = 1 $. Introducing the integrated resetting rate $ \Lambda_t = \int_0^t \lambda_s\, ds $, and solving Eq.~\ref{Eq:ODEFourier}, we obtain:
\begin{equation}
\hat P_r(k,t) = e^{-Dk^2 t - \Lambda_t} \left[ 1 + \int_0^t \lambda_s\, e^{Dk^2 s + \Lambda_s}\, ds \right].
\end{equation}

To further simplify this expression, we perform integration by parts on the integral term involving $ \lambda_s $, which yields:
\begin{equation}
\hat P_r(k,t) = 1 - Dk^2 \int_0^t e^{-Dk^2 (t - s) - (\Lambda_t - \Lambda_s)}\, ds.
\end{equation}
At this point, considering $ \lambda_t $ as a stochastic process, we are interested in computing the expected propagator $ \mathbb{E}[\hat P_r(k,t)] $ by averaging over all possible realizations of the resetting rate $ \lambda_t $:
\begin{equation}
\mathbb{E}[\hat P_r(k,t)] = 1 - Dk^2 \int_0^t e^{-Dk^2 (t - s)}\, \mathbb{E}\left[ e^{- (\Lambda_t - \Lambda_s)} \right] ds.
\end{equation}
The expectation $ \mathbb{E}[ e^{-(\Lambda_t - \Lambda_s)} ] $ encapsulates the impact of the stochastic resetting rate on the evolution of the propagator, linking the time-integrated variability of $ \lambda_t $ to the diffusion dynamics. Without specifying a particular stochastic time evolution law for $ \lambda_t $, this expression remains general, and any further calculation would require assumptions about the statistical characteristics of $ \lambda_t $ (such as its mean or autocorrelation).
Thus, the expected propagator for diffusion under a time-dependent and stochastic resetting rate is expressed as:
\begin{equation}
\mathbb{E}[\hat P_r(k,t)] = 1 - Dk^2 \int_0^t e^{-Dk^2 (t - s)}\, \mathbb{E}\left[ e^{- \int_s^t \lambda_u\, du } \right] ds,
\end{equation}
providing a form that is applicable for non-stationary and history-dependent rates. This formulation illustrates how the statistical properties of $ \lambda_t $, through the integrated quantity $ \Lambda_t $, shape the propagator’s evolution and ultimately influence the non-equilibrium steady state.

This derivation extends seamlessly to various diffusion processes by substituting $ Dk^2 $ with the general operator $ L(k) $, which characterizes the spatial dynamics specific to each diffusion model. Thus, the governing equation in Fourier space becomes:
\begin{equation}
\frac{d \hat P_r(k,t)}{d t} = -L(k) \hat P_r(k,t) - \lambda_t \hat P_r(k,t) + \lambda_t,    
\end{equation}
where $ L(k) $ is chosen according to the diffusion type under consideration. For classical diffusion, $ L(k) = Dk^2 $ describes the Gaussian spreading typical of heat conduction and Brownian motion. 

The fact that combining L\'evy-flight dynamics with resetting produces a first-order transition in the mean first-passage time as the jump exponent varies~\cite{kusmierz2014first} motivates the inclusion of the L\'evy case in our framework. Specifically, in anomalous diffusion processes with heavy-tailed displacements, such as L\'evy flights or fractional diffusion, the spatial operator takes the form $ L(k) = D_\alpha |k|^\alpha $, capturing the non-local jumps characteristic of such systems. These processes are particularly relevant in heterogeneous environments and ecological contexts. In convection-diffusion processes, where there is both diffusion and directional drift, the operator becomes $ L(k) = D k^2 + i \vec{v} \cdot \vec{k} $, which models systems like fluid flow where particles experience both random motion and drift. The additive resetting terms involving $ \lambda_t $ are independent of the choice of $ L(k) $, preserving the general structure of the solution: 
\begin{equation}
\mathbb{E}[\hat P_r(k,t)] = 1 - L(k) \int_0^t e^{-L(k) (t - s)}\, \mathbb{E}\left[ e^{- (\Lambda_t - \Lambda_s)} \right] ds.
\end{equation}

Consequently, this framework accommodates a broad spectrum of diffusive behaviors under a stochastic resetting rate, providing a unified approach for studying doubly stochastic resetting effects across local and non-local diffusion processes. This framework does not encompass space-dependent resetting rates~\cite{Roldan2017,Pinsky2018,Monthus2020}. It likewise does not encompass resetting on arbitrary network topologies, where resets may return to multiple nodes, though in that setting exact steady-state occupations and first-passage statistics have been obtained~\cite{gonzalez2021diffusive}.

\section{Long-time behaviour and non-equilibrium steady-state analysis}
We investigate the asymptotic behavior of the NESS for a diffusion process with a stochastic resetting rate $ \lambda_t $. The goal is to derive explicit expressions for the NESS distribution $ \hat P_r(k) $ in Fourier space and its inverse $ P_r(x) $ in real space. Starting with the expected propagator in Fourier space: 
\begin{equation}
\mathbb{E}[\hat P_r(k,t)] = 1 - Dk^2 \int_0^t e^{-Dk^2 (t - s)}\, \mathbb{E}\left[ e^{- \int_s^t \lambda_u\, du } \right] ds,
\end{equation}
we seek to analyze its long-time behavior as $ t \to \infty $. To facilitate this limit, we perform a change of variables by setting $ u = t - s $. This substitution allows us to rewrite the expectation in terms of the cumulative resetting rate $ \Lambda_u = \int_0^u \lambda_s \, ds $, which captures the integral of the rate over a time interval $ u $.
However, to express $ \mathbb{E}\left[ e^{-\int_s^t \lambda_u \, du } \right] $ as $ \mathbb{E}\left[ e^{-\Lambda_u} \right] $, we require $ \lambda_t $ to be a stationary stochastic process. This stationarity condition ensures that the statistical properties of $ \lambda_t $, and consequently of $ \Lambda_u $, depend only on the interval $ u = t - s $ rather than on the absolute times $ t $ and $ s $. With this assumption of stationarity, we can write:
\begin{equation}
\hat P_r(k) = \lim_{t \to \infty} \mathbb{E}[\hat P_r(k,t)] = 1 - Dk^2 \int_{0}^{\infty} e^{-Dk^2 u} \, \mathbb{E}\left[ e^{- \Lambda_u } \right] \, du,
\label{Eq:NESS}
\end{equation}
where $ D $ is the diffusion coefficient, and $ \mathbb{E}[e^{-\Lambda_u}] $ is the Laplace transform of the distribution of $ \Lambda_u $, incorporating the cumulative effect of the stochastic resetting rate over time. 

For the special case where $ \lambda_t = \lambda $ is constant, the cumulative resetting rate becomes $ \Lambda_u = \lambda u $, and we recover the NESS of diffusion with constant resetting rate ~\cite{evans2019resetting} (see supplementary information S1):
\begin{equation}
P_r(x) = \frac{1}{2}\sqrt{\frac{\lambda}{D}} e^{-\frac{\sqrt{\lambda}}{\sqrt{D}} |x|}.
\end{equation}
This exponential decay in $ P_r(x) $ defines a characteristic resetting length $ \ell_{\text{reset}} = \sqrt{\frac{D}{\lambda}} $, which quantifies the spatial extent over which diffusion spreads between resetting events. For distances $ |x| \gg \ell_{\text{reset}} $, the probability density decreases significantly, illustrating the localization effect induced by the constant resetting rate.

\begin{figure}[t]
    \centering
    \includegraphics[width=1\textwidth]{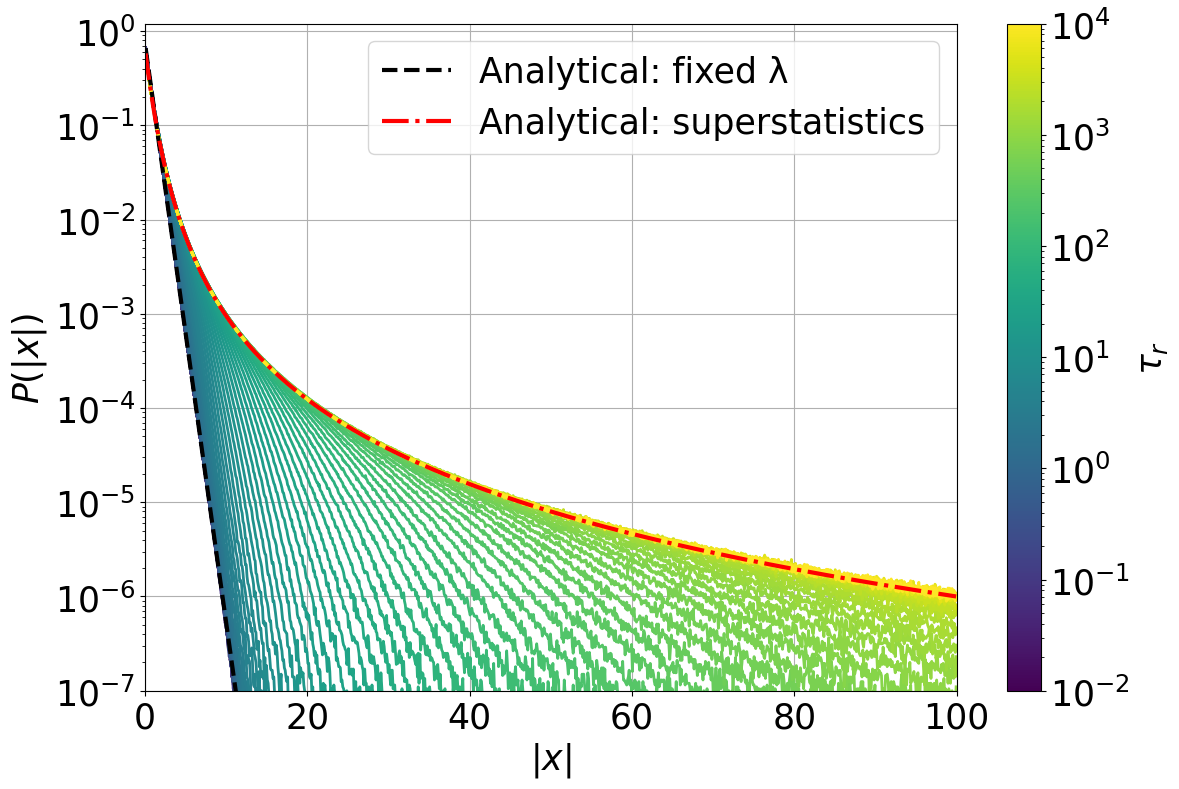}
        \caption{Probability density function $ P(|x|) $ for the absolute displacement of a one-dimensional diffusive particle subject to doubly stochastic resetting, computed numerically from simulations. The resetting rate is initially drawn from an exponential distribution with mean $\bar{\lambda} = 1.0$, and subsequently relaxes after a characteristic relaxation time $\tau_r$ to a fixed value $\bar{\lambda}$. Curves correspond to different values of $\tau_r$, color-coded according to the logarithmic scale indicated by the colorbar. Special analytical limits are shown for comparison: fixed-rate resetting ($\tau_r = 0$, black dashed line) and fully super-statistical resetting ($\tau_r = \infty$, red dot-dashed line). Simulation parameters: diffusion constant $D = 0.5$, number of independent trajectories $N_{\mathrm{sim}} = 10^{7}$.}
\label{fig:diffusive_resetting}
\end{figure}

Examining Eq.~\ref{Eq:NESS}, we see that the NESS distribution is controlled by the rate‐relaxation time $\tau_r$ through $\mathbb{E}[e^{-\Lambda_t}]$.  For early times $t\ll\tau_r$, the rate remains effectively frozen:
\begin{equation}
  \mathbb{E}[e^{-\Lambda_t}]
  \approx
  \mathbb{E}[e^{-\lambda t}]
  \equiv
  \tilde\lambda(t),\label{eq.ST}
\end{equation}
where we note that $\tilde\lambda(t)$ is the Laplace transform of the steady-state distribution of the fluctuating rate, evaluated at $t$.
For late times $t\gg\tau_r$, the rate self‐averages:
\begin{equation}
  \mathbb{E}[e^{-\Lambda_t}]
  \simeq
  \tilde\lambda(\tau_r)\,e^{-\bar\lambda\,(t-\tau_r)},
  \quad
  \bar\lambda=\mathbb{E}[\lambda]. \label{eq.LT}
\end{equation}
Splitting the integral in Eq.~\ref{Eq:NESS} at $u=\tau_r$, plugging in Eqs.~\ref{eq.ST} and \ref{eq.LT}, and evaluating each piece yields 
\begin{equation}
\hat P_r(k)
\simeq
\mathbb{E}\!\biggl[\frac{\lambda}{\lambda + D k^2}\biggr]
+ D k^2 \,
  \mathbb{E}\!\biggl[\frac{e^{-(D k^2 + \lambda)\,\tau_r}}{\lambda + D k^2}\biggr]
- \frac{D k^2 \, e^{-D k^2 \,\tau_r} \,\tilde\lambda(\tau_r)}
       {\bar\lambda + D k^2},
\label{eq:Prk_master}
\end{equation}
which in the limits $\tau_r\to0$ and $\tau_r\to\infty$ reduces to $\bar\lambda/(\bar\lambda + D k^2)$ and $\mathbb{E}[\lambda/(\lambda + D k^2)]$, respectively, illustrating the crossover between fixed‐rate and superstatistical behavior (see Fig.~\ref{fig:diffusive_resetting}).  Even when $\mathbb{E}[1/\lambda]$ diverges (e.g.\ for an exponential rate law), the small‐$k$ expansion involves only the finite combination $\mathbb{E}[(1-e^{-\lambda\tau_r})/\lambda]$, so all coefficients remain well‐defined.

In the limit $k\to0$, expanding (\ref{eq:Prk_master}) to first order in $D k^2$ gives
\begin{equation}
\hat P_r(k)\simeq
\begin{cases}
1 - \dfrac{D k^2}{\bar\lambda}\!\Bigl(1 + \tfrac12(\bar\lambda\tau_r)^2\Bigr),
& \bar\lambda\tau_r \ll 1,\\[8pt]
1 - \dfrac{D k^2}{\bar\lambda}\,\ln(\bar\lambda\tau_r),
& \bar\lambda\tau_r \gg 1,
\end{cases}
\end{equation}
so that in real space the tail remains exponential:

\begin{equation}
P_r(x)\sim e^{-\sqrt{\frac{\lambda_{\mathrm{eff}}}{D}}|x|},
  \quad
  \lambda_{\mathrm{eff}}\simeq
  \begin{cases}
  \dfrac{\bar\lambda}{1 + \tfrac12(\bar\lambda\tau_r)^2}, & \bar\lambda\tau_r\ll1,\\[8pt]
  \dfrac{\bar\lambda}{\ln(\bar\lambda\tau_r)},              & \bar\lambda\tau_r\gg1.
  \end{cases}
\label{eq:lambda_eff}
\end{equation}
For $k\to\infty$ the first term of (\ref{eq:Prk_master}) dominates, $\hat P_r(k)\sim\frac{\bar\lambda}{Dk^{2}}$, implying a Laplace core $P_r(x)\simeq P_r(0)e^{-r|x|}$ with $r=\frac{\bar\lambda}{2DP_r(0)}$.  Integration of (\ref{eq:Prk_master}) over $k$ yields:
\begin{equation}
    P_r(0) = \frac{1}{2}\mathbb{E}\left[ \sqrt{\frac{\lambda}{D}}\text{erf}\left(\sqrt{\lambda\tau_r}\right)\right]+\frac{1}{2}\sqrt{\frac{\bar\lambda}{D}}e^{\bar\lambda\tau_r}\tilde\lambda(\tau_r)\text{erfc}\left(\sqrt{\bar\lambda\tau_r}\right)
\end{equation}
We can expand the exact results in the relevant asymptotic regimes:
\begin{equation}
P_r(0)\simeq
\begin{cases}
\dfrac12\sqrt{\frac{\bar\lambda}{D}}\!\left(1-\dfrac{1}{6\sqrt{\pi\bar\lambda}}\sigma^{2}\tau_r^{3/2}\right), & \bar\lambda\tau_r\ll1,\\
\dfrac{\sqrt{\pi}}{4}\sqrt{\frac{\bar\lambda}{D}}\!\left(1+\dfrac{1}{\sqrt{\pi}}(\bar\lambda\tau_r)^{-3/2}\right), & \bar\lambda\tau_r\gg1,
\end{cases}\qquad
\sigma^{2}=\mathbb{E}\left[ \lambda^{2}\right]-\bar\lambda^{2},
\end{equation}
and therefore
\begin{equation}
r(\tau_r)\simeq
\begin{cases}
\sqrt{\frac{\bar\lambda}{D}}\!\left(1+\dfrac{1}{6\sqrt{\pi}}
\dfrac{\sigma^{2}}{\sqrt{\bar\lambda}}\tau_r^{3/2}\right), & \bar\lambda\tau_r\ll1,\\
\dfrac{2}{\sqrt{\pi}}\sqrt{\frac{\bar\lambda}{D}}\!\left(1-\dfrac{2}{\pi}(\bar\lambda\tau_r)^{-3/2}\right), & \bar\lambda\tau_r\gg1.
\end{cases}
\label{eq:r_tau}
\end{equation}

When $\tau_r\to\infty$ the rate remains frozen for each trajectory and the ensemble‑averaged density is the super‑statistical mixture
\begin{equation}
P_r(x)=\mathbb{E}\left[ \tfrac12\sqrt{\frac{\lambda}{D}}\;
           e^{-\sqrt{\frac{\lambda}{D}}\,|x|}\right],
\label{eq:superstat_mix}
\end{equation}
first recognized in the context of driven non-equilibrium systems \cite{beck2003superstatistics}.  For an exponential distribution $f_\lambda(\lambda)=\bar\lambda^{-1}e^{-\lambda/\bar\lambda}$ the integral in (\ref{eq:superstat_mix}) can be performed exactly,
\begin{equation}
P_r(x)
=
\frac{\sqrt\pi\,\sqrt{\bar\lambda}}{8\,D^{3/2}}\,
\bigl(\bar\lambda\,x^2 + 2D\bigr)\,
\exp\!\left(\tfrac{\bar\lambda\,x^2}{4D}\right)\,
\text{erfc}\!\left(\tfrac{|x|}{2}\sqrt{\tfrac{\bar\lambda}{D}}\right)
-\frac{\bar\lambda}{4D}\,\lvert x\rvert.
\label{eq:P_exact}
\end{equation}
which approaches $\frac{\sqrt{\pi}}{4}\sqrt{\frac{\bar\lambda}{D}}$ at the origin and crosses over to the algebraic tail $P_r(x)\sim \frac{2D}{\bar\lambda}|x|^{-3}$ as $|x|\to\infty$. The power‑law regime is the spatial signature of trajectories that spend long epochs in anomalously small resetting rates, a mechanism analogous to the non‑Gaussian cores observed for Brownian motion in fluctuating diffusivity landscapes \cite{chechkin2017brownian}.

Equations (\ref{eq:Prk_master})–(\ref{eq:P_exact}) summarize how a finite $\tau_r$ tunes the crossover from the classical self‑averaging exponential steady state to a disorder‑dominated, super‑statistical regime: fast rate relaxation leaves both core and tail essentially unchanged up to small $(\tau_r)^{3/2}$ corrections, whereas slow relaxation logarithmically broadens the exploration length and softens the central peak, providing a quantitative baseline for experiments where the resetting protocol can itself fluctuate in time. Physically, this crossover arises from two competing length scales. First, the diffusive‐exploration length $\ell_D =\sqrt{D\tau_r}$ is the typical distance a particle explores during one rate‐plateau of duration $\tau_r$. Second, the resetting length $\ell_{\rm reset} \;=\;\sqrt{\frac{D}{\bar\lambda}}$ sets the scale over which resets dominate diffusion. For $|x|\ll\ell_D$, the rate is effectively frozen and each trajectory samples a super‐statistical mixture of exponentials $e^{-\sqrt{\lambda/D}\,|x|}$. In contrast, for $|x|\gg\ell_D$, many rate‐fluctuation intervals intervene and self‐average the tail into a single exponential $e^{-\lambda_{\rm eff}|x|}$ with $\lambda_{\rm eff}$ from \eqref{eq:lambda_eff}. In the limit $\tau_r\to\infty$, rare, extended low‐$\lambda$ plateaus generate an algebraic $|x|^{-3}$ tail, marking the dominance of long excursions in weak‐reset environments.

\section{Exactly solvable non-equilibrium steady state}
\begin{figure}[htbp]
    \centering
    \includegraphics[width=\textwidth]{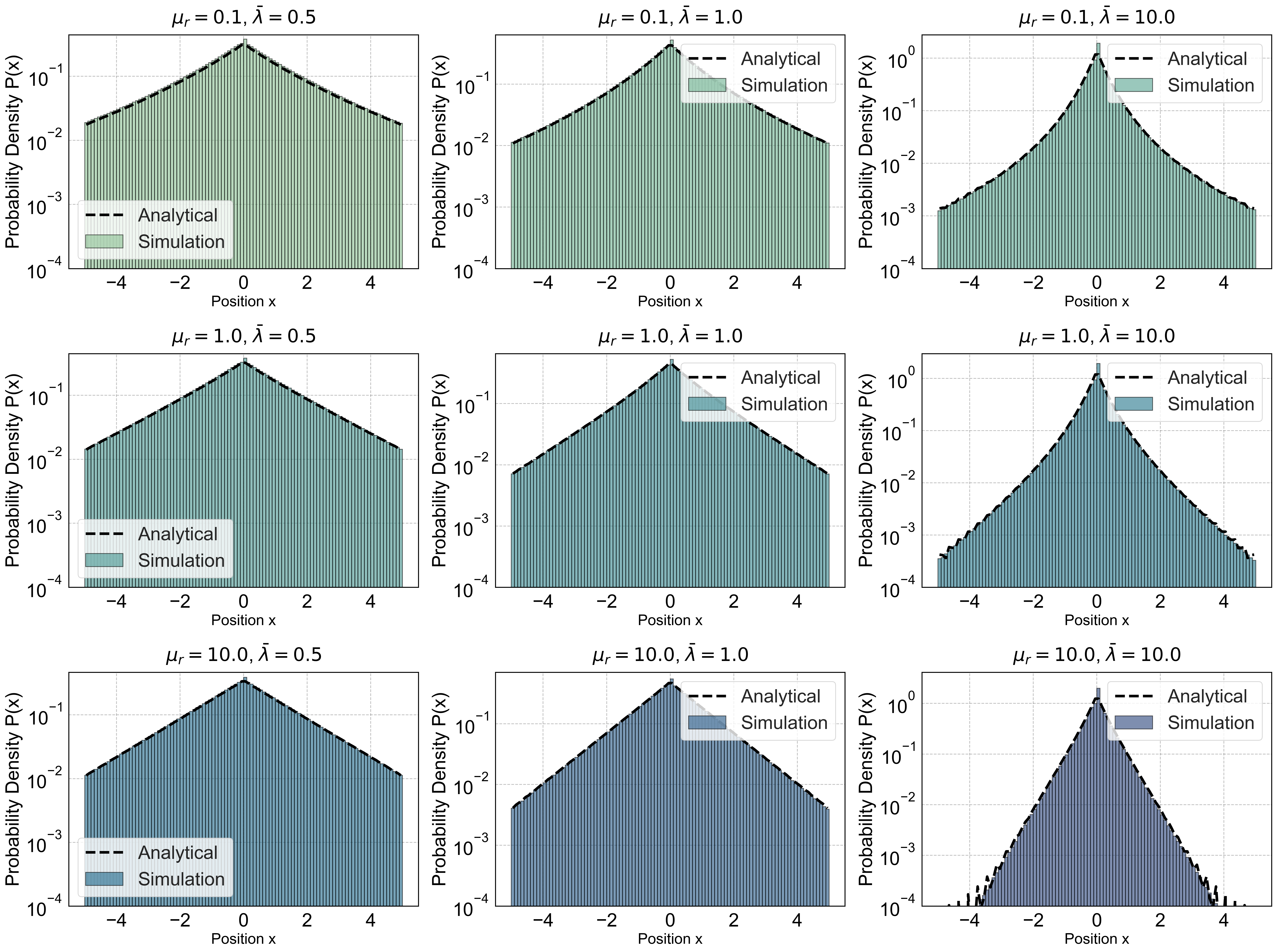}
    \caption{
        Comparison of simulated and analytical probability densities $ P(x) $ for the exactly solvable doubly stochastic resetting process. Each subplot displays the probability density across position $ x $, computed using both simulation and analytical approaches for different values of the mean resetting rate $ \mu_r $ and the mean resetting intensity $ \bar\lambda $. The simulation results were obtained by running 10,000 realizations of a diffusion process with a diffusion coefficient $ D = 1.0 $, a total simulation time of $ t_{\text{max}} = 100 $, a time step of $ \Delta t = 0.01 $, plotted on a spatial range $ x \in [-5, 5] $ divided into 100 bins. In the simulations, the resetting rate $ \lambda_t $ is sampled from an exponential distribution with mean $ \bar \lambda $, and it changes randomly at an average rate $ \mu_r $. The diffusion process undergoes resetting to the origin with probability $ \lambda_t \Delta t $ at each time step. For each subplot, the simulated probability density (represented by the bar histograms) is shown alongside the analytical solution (black dashed line), computed through numerical inversion of $ \hat P_r(k) $. The panels are organized as follows: the top, middle, and bottom rows correspond to mean resetting rates $ \mu_r = 0.1 $, $ \mu_r = 1 $, and $ \mu_r = 10 $, respectively. The columns from left to right represent mean resetting intensities $ \bar\lambda = 0.5 $, $ \bar\lambda = 1 $, and $ \bar\lambda = 10 $. Each subplot title indicates the specific values of $ \mu_r $ and $ \bar\lambda $ used. The y-axis is plotted on a logarithmic scale.
    }
    \label{fig:doubly_stochastic_resetting}
\end{figure}

A particularly tractable model of doubly‑stochastic resetting is obtained when the rate itself is quenched for random plateaus of duration $\tau$ drawn from the exponential law $p_\tau(\tau)=\mu_r e^{-\mu_r\tau}$, with $\mu_r=\tau_r^{-1}$ the inverse of the relaxation time of the rate defined in the previous model, while on each plateau the value of the rate, denoted $\lambda$, is redrawn independently from a prescribed distribution $f_\lambda(\lambda)$. During a plateau the particle undergoes standard diffusion with coefficient $D$ and is reset to the origin with the frozen rate $\lambda$. Renewal theory yields an exact expression for the stationary propagator in Fourier space (see supplementary information S2),
\begin{equation}
\hat P_r(k)=
\frac{1-\left(\mu_r+Dk^{2}\right)\,\tilde{\tilde{f_\lambda}}\!\left(\mu_r+Dk^{2}\right)}
     {1-\mu_r\,\tilde{\tilde{f_\lambda}}\!\left(\mu_r+Dk^{2}\right)},
\qquad
\tilde{\tilde{f_\lambda}}(u)=\int_0^\infty\frac{f_\lambda(\lambda)}{u+\lambda}\,d\lambda,
\label{eq:PRk_general}
\end{equation}
which is valid for any choice of $f_\lambda(\lambda)$.  Two natural length scales govern the ensuing steady state: the diffusive distance explored in one rate‑refresh time, $\ell_D=\sqrt{D/\mu_r}$, and the mean inter‑reset distance, $\ell_\text{reset}=\sqrt{D/\bar\lambda}$ .  Their ratio $\gamma=\sqrt{\bar\lambda/\mu_r}$ organizes the physics. When $\gamma\ll1$ the rate is reshuffled many times before a typical reset occurs; in this self‑averaging limit $\tilde f_\lambda(u)\simeq\frac{1}{u}-\frac{\bar\lambda}{u^2}$ and (\ref{eq:PRk_general}) collapses to $\hat P_r(k)=\frac{\bar\lambda}{\bar\lambda+Dk^{2}}$, giving the usual exponential profile $P_r(x)\sim\exp\left(-\sqrt{\frac{\bar\lambda}{D}}\,|x|\right)$.  In the opposite limit $\gamma\gg1$ each trajectory keeps a fixed rate for many resets: averaging must then be performed after evolving, and (\ref{eq:PRk_general}) reduces to the super‑statistical mixture
$\hat P_r(k)=\mathbb{E}\left[ \frac{\lambda}{\lambda+Dk^{2}}\right]$ whose inverse transform decays algebraically, $P_r(x)\sim\frac{2D}{\bar\lambda}|x|^{-3}$, reflecting the dominance of exceptionally small‑$\lambda$ plateaus at large distances.

The experimentally relevant case of an exponential distribution $f_\lambda(\lambda)=\bar\lambda^{-1}e^{-\lambda/\bar\lambda}$ makes the crossover explicit: inserting this law into (\ref{eq:PRk_general}) yields

\begin{equation}
\hat P_r(k)=
\frac{1-\dfrac{Dk^{2}+\mu_r}{\bar{\lambda}}\,
      E_1\!\left(\frac{Dk^{2}+\mu_r}{\bar{\lambda}}\right)\,
      e^{\frac{Dk^{2}+\mu_r}{\bar{\lambda}}}}
     {1-\dfrac{\mu_r}{\bar{\lambda}}\,
      E_1\!\left(\frac{Dk^{2}+\mu_r}{\bar{\lambda}}\right)\,
      e^{\frac{Dk^{2}+\mu_r}{\bar{\lambda}}}},
\label{eq:PRk_exp}
\end{equation}
with $E_{1}(z)=\int_{z}^{\infty}\!u^{-1}e^{-u}\,du$ the exponential integral.  Expanding (\ref{eq:PRk_exp}) at small $k$ one finds an exponential tail $P_r(x)\propto\exp[-\lambda_{\rm eff}|x|]$ whose decay constant interpolates smoothly between $\lambda_{\rm eff}=\bar\lambda$ for $\gamma\ll1$ and $\lambda_{\rm eff}=\bar\lambda/\ln\gamma^{2}$ for $\gamma\gg1$.  In the bulk region $|x|\lesssim \ell_D=\sqrt{D/\mu_r}$ the profile remains approximately exponential with slope $\sqrt{\frac{\bar\lambda}{D}}$ in the fast‑refresh regime and $\frac{2}{\sqrt\pi}\sqrt{\frac{\bar\lambda}{D}}$ in the slow‑refresh regime, the prefactor difference arises from the excess probability mass $P_r(0)$ accumulated at the origin when rate plateaus become long. Equation (\ref{eq:PRk_exp}) therefore provides a compact, experimentally testable bridge between the annealed‑rate and quenched‑rate limits, controlled by the single dimensionless parameter $\gamma$. As confirmed in Fig.~\ref{fig:doubly_stochastic_resetting}, the simulation histograms and analytical curves (dashed lines) for nine distinct $(\mu_r,\bar\lambda)$ pairs faithfully reproduce the two limiting forms, narrow exponential for $\gamma\ll1$ and algebraic $|x|^{-3}$ tail for $\gamma\gg1$, while intermediate values of $\gamma=\sqrt{\bar\lambda/\mu_r}$ generate the smooth interpolation in both the bulk slope and the far‐tail decay anticipated by Eq.~(\eqref{eq:PRk_exp}). Because $\mu_r$ can be tuned independently of $\bar\lambda$, for instance via feedback in optical‑trap implementations, one may smoothly steer the system from standard exponentially localized resetting to a broad, power‑law‑tailed localization regime, and thereby quantify the impact of temporal disorder on reset‑driven search processes.

\section{Conclusion}
We have investigated diffusion processes subject to stochastic resetting where the resetting rate $\lambda_t$ itself fluctuates randomly over time, leading to a doubly stochastic framework. This approach extends traditional models by considering a resetting rate that evolves independently and is not renewed upon each return to the origin. Our analysis reveals that such fluctuations in the resetting rate impact the non-equilibrium steady state of the system, introducing new structural features in the PDF. 

At short distances, the system exhibits super-statistical behavior due to fluctuations in the resetting rate. These fluctuations dominate the dynamics, resulting in non-trivial distributions that deviate from those predicted by models with constant or simply time-dependent resetting rates. The local PDF reflects a mixture of exponential decays corresponding to the various instantaneous resetting rates experienced by the system. This finding underscores the significant influence of rate fluctuations on the microscopic structure of the steady state, highlighting the importance of considering the full statistical properties of $\lambda_t$.

As we consider longer distances, the cumulative effect of the fluctuating resetting rate over extended timescales leads to an averaging out of these fluctuations. The PDF transitions to a form governed by an effective resetting rate, which may differ notably from the mean resetting rate depending on the relaxation dynamics of $\lambda_t$. Specifically, when the relaxation timescale $\tau_r$ of the resetting rate is much smaller than the typical resetting time, the effective rate converges to the mean rate due to swift self-averaging. Conversely, when $\tau_r$ is large, there is a pronounced bias toward lower resetting rates over long distances, resulting in an effective rate significantly lower than the mean. This yields a broader tail in the PDF, indicating enhanced spatial exploration by the diffusing particles. This dual structure emphasizes the crucial roles of both fluctuations and long-term average behaviors of the resetting rate in shaping the NESS.

The phenomenology we observe proves particularly relevant in physical systems where the resetting mechanism is subject to environmental fluctuations or intrinsic noise. For example, enzymatic turnovers can be mapped onto our resetting picture by viewing the enzyme–substrate complex as diffusing along a reaction coordinate $x$, with product release corresponding to a reset to $x=0$.  If the instantaneous catalytic rate $k_{\rm cat}(t)$ fluctuates, due to temperature drifts, pH shifts, or transient inhibitor binding, then setting $\lambda_t=k_{\rm cat}(t)$ makes each turnover a doubly stochastic reset. Our framework then delivers closed‐form predictions for both the distribution of turnover intervals and the steady‐state distribution of pre‐turnover excursions: in the rapid‐fluctuation limit one recovers the familiar exponential kinetics and localized steady state. Whereas slow or large‐amplitude rate fluctuations give rise to a three‐regime structure, a super‐statistical core of short excursions, a power‐law regime at intermediate scales, and an asymptotic exponential tail, thereby quantifying exactly how environmental noise broadens and skews enzyme‐catalyzed turnover‐time and concentration profiles beyond the constant‐rate paradigm. Similarly, in search processes or foraging behavior, the rate at which a searcher resets its search strategy or returns to a starting point may vary due to fatigue, environmental factors, or adaptive behaviors. Our findings provide insights into how these variations affect the efficiency and patterns of search strategies. Voltage‐gated ion channels in neurons, whose open and closed‐state transitions depend on membrane potential noise and feedback, can be recast as doubly stochastic resets, explaining the broad distribution of channel dwell times. Recent feedback‑controlled optical trap experiments~\cite{talfriedman2020experimental} could in principle impose a stochastic reset rate $\lambda_t$ and thus test the crossover reported here. Indeed, an optical‐tweezer study has measured the optimal reset rate and observed a metastable transition in mean search times~\cite{besga2020optimal}. Finally, polymer translocation through nanopores, where the attempt rate depends on polymer conformation and pore interactions, naturally exhibits heavy‐tailed translocation times when the reset rate is treated as a fluctuating variable. These examples underscore the wide applicability of our model to systems in which the reset mechanism itself is subject to environmental or intrinsic noise.

An important extension of our work would be to examine how other observables are affected by the stochasticity of the resetting rate. Specifically, the first-passage time (FPT) to a target is a crucial metric in many processes, such as reaction kinetics, transport phenomena, and search problems. Investigating how the distribution of FPT is altered by a fluctuating resetting rate would provide a deeper understanding of the temporal aspects of such systems. We anticipate that variability in the resetting rate could lead to non-exponential survival probabilities and affect the mean and variance of the FPT, potentially enhancing or impeding the likelihood of reaching the target depending on the characteristics of $\lambda_t$. 

Our analysis can also be extended to consider different types of stochastic processes for the resetting rate. For example, incorporating temporal correlations in $\lambda_t$ or modeling it as a Hawkes process could reveal additional complexities in the NESS. Such extensions would be directly building on our current framework and could capture memory effects present in real systems where fluctuations are not entirely random but exhibit some degree of temporal correlation. It would also be intriguing to ask whether the introduction of a stochastic rate has implications for the energetic cost of resetting. Recent studies have begun to quantify the work required to maintain resetting protocols~\cite{olsen2024thermodynamic}. Incorporating a fluctuating rate into such analyses could reveal new optimal strategies or energetic signatures of doubly stochastic resetting.

In conclusion, our study demonstrates that fluctuations in the resetting rate significantly influence the non-equilibrium steady states of diffusion processes. By revealing the scaling regimes in the PDF and quantifying the impact of rate fluctuations, we provide a deeper understanding of how temporal variability in resetting mechanisms alters the spatial distribution of diffusing particles under resetting. These insights have potential applications in modeling systems where the resetting dynamics are inherently stochastic.

\section*{References}

\providecommand{\newblock}{}

\end{document}